\documentstyle[prc,aps,epsfig,eqsecnum]{revtex}

\begin{document}
\draft
\title{alpha-nucleus potentials for the neutron-deficient $p$ nuclei}
\author{P.~Mohr}
\address{
  Institut f\"ur Kernphysik, Technische Universit\"at Darmstadt,
  Schlossgartenstrasse 9, D--64289 Darmstadt, Germany
}
\date{\today}
\maketitle
\begin{abstract}
$\alpha$-nucleus potentials are one important ingredient for the
understanding of the nucleosynthesis of heavy neutron-deficient $p$ nuclei
in the astrophysical $\gamma$-process where these $p$ nuclei are produced
by a series of ($\gamma$,n), ($\gamma$,p), and ($\gamma$,$\alpha$) reactions.
I present an improved $\alpha$-nucleus potential at the astrophysically
relevant sub-Coulomb energies which is derived from the analysis of
$\alpha$ decay data and from a previously established systematic
behavior of double-folding potentials.
\end{abstract}

\pacs{PACS numbers: 23.60.+e,25.55.Ci,26.30+k,26.45.+h}



\section{Introduction}
\label{sec:intro}
The bulk of the heavy nuclei ($A \ge 100$) has been synthesized
by neutron capture in the $s$- and $r$-process. However, most of the rare
neutron-deficient nuclei with $A \ge 100$ 
cannot be produced by neutron capture.
The main production mechanism for these so-called $p$ nuclei
is photodisintegration in the astrophysical $\gamma$-process 
by
($\gamma$,n), 
($\gamma$,p), and
($\gamma$,$\alpha$) reactions
of heavy seed nuclei from the
$s$- and $r$-process. 
A list of the neutron-deficient $p$ nuclei from $^{74}$Se to
$^{196}$Hg can be found in Table 1 of Ref.~\cite{Lam92}.
Typical parameters for the $\gamma$-process are temperatures of
$2 \le T_9 \le 3$ ($T_9$ is the temperature in GK), 
densities of about $10^6$\,g/cm$^3$, and time
scales in the order of one second. Several astrophysical sites 
for the $\gamma$-process have been proposed, 
and the oxygen- and neon-rich layers of type II supernovae
seem to be a good candidate. However, there has been no
definite conclusion reached yet
with respect to the astrophysical site
where the $\gamma$-process occurs.
Details about the astrophysical scenarios
can be found in the reviews by 
Lambert \cite{Lam92}, 
Arnould and Takahashi \cite{Arn99},
Wallerstein {\it et al.}~\cite{Wal97},
and in Refs.~\cite{Ito61,Woo78,Ray90,Pra90,How91}.

Almost no experimental data exist
for the cross sections 
of the $\gamma$-induced reactions
at astrophysically relevant energies. 
Therefore, all reaction rates have been derived theoretically
using statistical model calculations.
One striking example is $^{146}$Sm
which is a
potential chronometer for the $\gamma$-process.
The production ratio of $^{146}$Sm and $^{144}$Sm
depends sensitively on the ($\gamma$,n) and ($\gamma$,$\alpha$) 
cross sections at the branching nucleus $^{148}$Gd,
and it has been shown that especially the ($\gamma$,$\alpha$) cross section
can be calculated only if a reliable $\alpha$-nucleus potential
is available. Predictions from different potentials
differ by one order of magnitude \cite{Woos90,Rau95,Mohr97,Som98}.
The need for improved $\alpha$-nucleus potentials for astrophysical
calculations has been pointed out in \cite{Rau98,Gra98}.

Systematic $\alpha$-nucleus potentials have been presented in several papers
(e.g.\ Refs.\ \cite{Mann78,Av94,Atz96}), and recently these studies have been
extended to astrophysically relevant energies
\cite{Rau98,Gra98}. Usually potentials are derived from scattering data.
However, at the astrophysically relevant energies below the Coulomb
barrier it is difficult to derive the potential from the experimental
data unambiguously (see e.g.\ \cite{Mohr97}). Alternatively, the cluster
model provides another possibility to determine $\alpha$-nucleus potentials
by an adjustment of the $\alpha$-nucleus potential to the bound state
properties of the nucleus $(A+4) = A \otimes \alpha$
\cite{Mich77,Abele93,Buck95}. The half-lives of
many $\alpha$ emitters have been calculated in \cite{Buck93,Buck96}
using a specially shaped potential,
but this potential had to be modified 
to describe elastic scattering for $^{208}$Pb \cite{Buck96a}.
It has been shown in Refs.~\cite{Hoy94,Atz96} that a systematic
double-folding potential is able to reproduce both the bound state
properties of $^{212}$Po = $^{208}$Pb $\otimes$ $\alpha$ and
elastic $\alpha$ scattering of $^{208}$Pb. Finally,
folding potentials give an excellent description of the 
experimental data
on $^{144}$Sm at 20\,MeV \cite{Mohr97}, but again 
the potential which was used for the calculation of the $\alpha$ decay 
data \cite{Buck93} is not able to reproduce the precision scattering data.

The basic idea of this work is to determine the $\alpha$-nucleus potential
at astrophysically relevant energies of about 5 to 12\,MeV which is 
in the energy gap between the bound state potentials and the scattering
potentials. I have chosen double folding potentials because of the small
numbers of adjustable parameters. A systematic behavior of the strength
of double folding potentials at higher energies
is already given in Ref.~\cite{Atz96}. In
this work I will analyze bound state properties. First, I will briefly present
the method in Sect.~\ref{sec:meth}, then I will give results for the 
neutron-deficient $\alpha$ emitting $p$ nuclei in Sect.~\ref{sec:res},
and finally I will give an outlook on possible further improvements
of the $\alpha$-nucleus potentials (Sect.~\ref{sec:out}).

\section{Folding potentials and $\alpha$ decay}
\label{sec:meth}
The $\alpha$ decay of nuclei is a clear proof that the wave function
of the $\alpha$ emitting nucleus $A+4$ has a non-negligible component
$A \otimes \alpha$, whereas low-energy elastic scattering is
described using a pure $A \otimes \alpha$ wave function. 
Therefore, an effective $\alpha$-nucleus potential should describe
simultaneously the half-life of the $\alpha$ emitter $A+4$ and
elastic $\alpha$ scattering of the nucleus $A$. In the 
astrophysically relevant
energy region the $\alpha$-nucleus potential is accessible mainly
from the decay data because elastic scattering is dominated by the 
Coulomb interaction. For the analysis of the $\alpha$ decay data
I apply the semi-classical model of Ref.~\cite{Gur87}, and the
nuclear potentials $V_N(r)$ are calculated by the double-folding procedure:
\begin{equation}
V_N(r) = \lambda V_F(r) = 
  \lambda \int \int \rho_P(r_P) \, \rho_T(r_T) \,
  v_{\rm{eff}}(E,\rho = \rho_P + \rho_T, s = |\vec{r}+\vec{r_P}-\vec{r_T}|) \,
  d^3r_P \, d^3r_T
\label{eq:fold}
\end{equation}
where $\rho_P$, $\rho_T$ are the densities of projectile resp.~target,
and $v_{\rm{eff}}$ is the effective nucleon-nucleon interaction taken in the
well-established DDM3Y parametrization \cite{Sat79,Kob84}.
Details about the folding procedure can be found in 
Refs.~\cite{Abele93,Atz96}, the folding integral (\ref{eq:fold})
has been calculated using the code DFOLD \cite{DFOLD}.
The strength of the folding potential is adjusted by the usual
strength parameter $\lambda$ with $\lambda \approx 1.1 - 1.3$
leading to volume integrals $J_R$ per interacting nucleon pair
of about 300 to 350\,MeV\,fm$^3$.
$J_R$ is defined by
\begin{equation}
J_R = \frac{4\pi}{A_P A_T} \int_0^\infty V_N(r) \, r^2 \, dr \quad \quad .
\label{eq:vol}
\end{equation}
Note that in the discussion of volume integrals $J$ usually the negative sign
is neglected; also in this paper all $J$ values are negative.

The densities of the nuclei
have been derived from the experimentally known charge
density distributions \cite{devries87} and assuming identical
proton and neutron distributions. For nuclei where no experimental
charge density distribution is available 
(i) the density distribution
of the closest neighboring isotope was used with an adjusted
radius parameter $R \sim A^{1/3}$
($^{186}$Os, $^{182}$W, $^{170}$Yb, $^{150}$Gd),
(ii) the average between two neighboring stable isotopes was used
($^{146}$Sm) and
(iii) the average of $^{138}$Ba and $^{142}$Nd was used for
$^{140}$Ce.

The total potential is given by the sum of the nuclear
potential $V_N(r)$ and the Coulomb potential $V_C(r)$:
\begin{equation}
V(r) = V_N(r) + V_C(r) \quad \quad .
\label{eq:vtot}
\end{equation}
The Coulomb potential is taken in the usual form of a 
homogeneously charged sphere where the Coulomb radius $R_C$
has been chosen identically with the $rms$ radius of the
folding potential $V_F$.

The potential strength parameter $\lambda$ was adjusted to the
energy of the $\alpha$ particle in the $\alpha$ emitter
$(A+4) = A \otimes \alpha$. The number of nodes of the bound state
wave function was taken from the Wildermuth condition
\begin{equation}
Q = 2N + L = \sum_{i=1}^4 (2n_i + l_i) = \sum_{i=1}^4 q_i
\label{eq:wild}
\end{equation}
where $Q$ is the number of oscillator quanta,
$N$ is the number of nodes and $L$ the relative angular
momentum of the $\alpha$-core wave function, and
$q_i = 2n_i + l_i$ are the corresponding quantum numbers
of the nucleons in the $\alpha$ cluster. I have taken
$q = 4$ for $50 < Z,N \le 82$,
$q = 5$ for $82 < Z,N \le 126$, and
$q = 6$ for $N > 126$
where $Z$ and $N$ are the proton and neutron number of the daughter nucleus
(see also Table \ref{tab:res}).

The $\alpha$ decay width $\Gamma_\alpha$ is given by the following
formulae \cite{Gur87}:
\begin{equation}
\Gamma_\alpha = P F \frac{\hbar^2}{4\mu} 
\exp{\left[ -2 \int_{r_2}^{r_3} k(r) dr \right]}
\label{eq:gamma}
\end{equation}
with the preformation factor $P$, the normalization factor $F$ 
\begin{equation}
F \int_{r_1}^{r_2} \frac{dr}{k(r)} = 1
\label{eq:f}
\end{equation}
and the wave number $k(r)$
\begin{equation}
k(r) = \sqrt{ \frac{2\mu}{\hbar^2}\left|E - V(r)\right|} \quad \quad .
\label{eq:k}
\end{equation}
$\mu$ is the reduced mass and $E$ is the
decay energy of the $\alpha$ decay which was taken from the computer files
based on the mass table of Ref.~\cite{Audi95}. 
The $r_i$ are the classical turning points. For $0^+ \rightarrow 0^+$
$s$-wave decay the inner turning point is at $r_1 = 0$. $r_2$ varies
from about 7 to 9\,fm, and $r_3$ varies from 45 up to about 90\,fm.
The decay width $\Gamma_\alpha$ is related
to the half-life by the well-known relation 
$\Gamma_\alpha = \hbar \ln{2} / T_{1/2}$.

It has to be pointed out that the preformation
factor $P$ should be smaller than unity because the simple two-body
model assumes that the ground state wave function of the $\alpha$ emitter
$A+4$ contains a pure $A \otimes \alpha$ configuration. The decay width
in this model therefore always overestimates the experimental decay width.
I determine the preformation factor $P$ from the ratio between the
calculated and the experimental half-lives \cite{NNDC}.
A strong $A \otimes \alpha$ cluster component is expected for nuclei $A$
with magic proton and/or neutron numbers, and indeed the calculations
show increased values for $P$ around $N = 126$ ($^{208}$Pb) and $N = 82$
(e.g.\ $^{140}$Ce) (see Table \ref{tab:res}). 

\begin{figure}
\begin{center}
\epsfig{figure=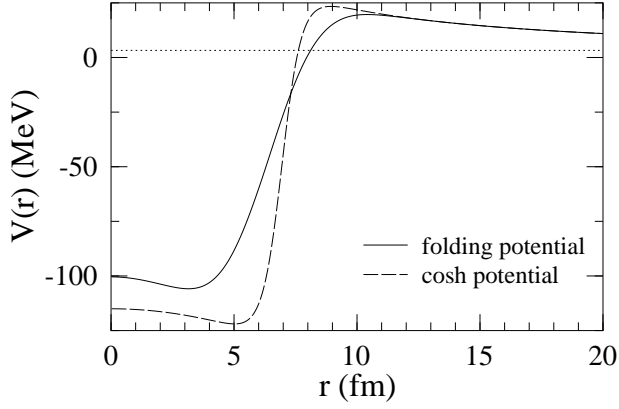,bbllx=60,bblly=60,bburx=580,bbury=350,
width=10.0cm}
\end{center}
\caption{
  \label{fig:pot} 
  Comparison of the potentials $V(r) = V_N(r) + V_C(r)$
  from this work (folding potential, full line) and from
  \protect\cite{Buck93} (specially shaped $\cosh$ potential, dashed line)
  for the system $^{190}$Pt = $^{186}$Os $\otimes$ $\alpha$.
  The decay energy $E$ is indicated by a dotted line.
  Note the significantly higher Coulomb barrier in \protect\cite{Buck93}
  compared to the folding potential.
}
\end{figure}

A preformation factor of $P = 1$ as used in
\cite{Buck93} seems to be the consequence 
of the specially shaped $\cosh$ potential of that work.
As an example I compare the potentials $V(r) = V_N(r) + V_C(r)$ 
from this work and from \cite{Buck93}
for the system $^{190}$Pt = $^{186}$Os $\otimes$ $\alpha$
in Fig.~\ref{fig:pot}.
The $rms$ radius of the potential from \cite{Buck93} is significantly
smaller ($r_{rms} = 5.58$\,fm) than the $rms$ radius of the folding
potential ($r_{rms} = 5.97$\,fm). Therefore, the Coulomb
barrier in \cite{Buck93} is significantly higher and 
the calculated half-lives in \cite{Buck93} 
are roughly one order of magnitude larger
than in this work.
Note that in \cite{Buck93} the potential was adjusted only to
decay properties with the assumption $P = 1$ whereas in this
work an effective potential is presented which is designed to 
describe decay properties and scattering wave functions
and which leads to realistic preformation factors $P$.

\section{Results for neutron-deficient $\alpha$ emitters}
\label{sec:res}
\begin{figure}
\begin{center}
\epsfig{figure=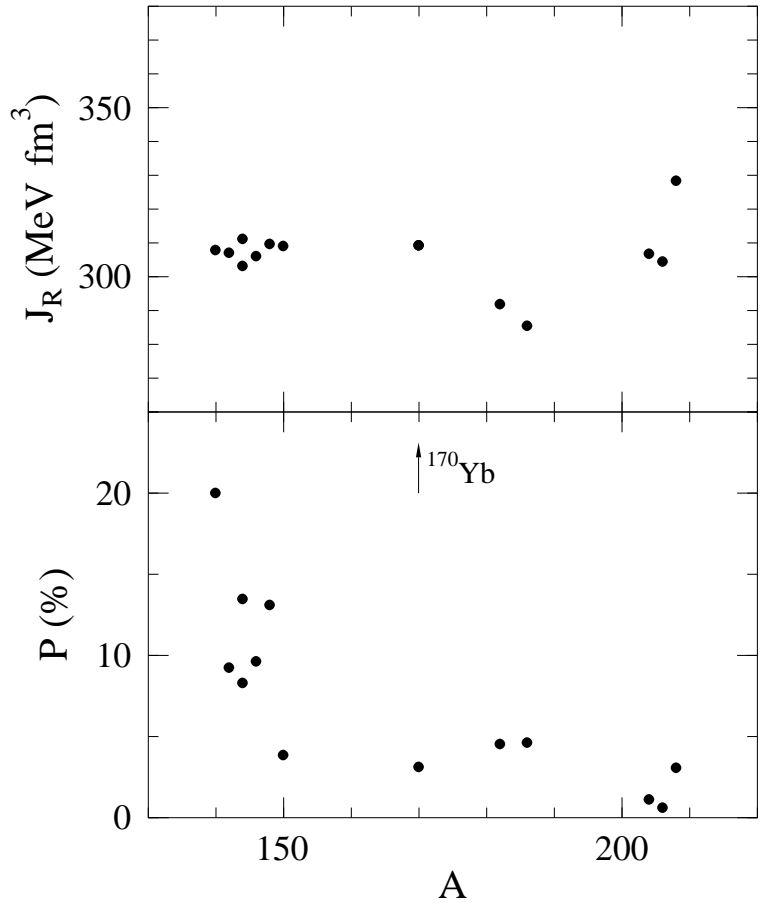,bbllx=60,bblly=60,bburx=480,bbury=580,
width=10.0cm}
\end{center}
\caption{
  \label{fig:res} 
  The volume integrals $J_R$ (upper diagram)
  and the preformation factors $P$ (lower diagram)
  are shown for several neutron-deficient $\alpha$ emitters.
  The variation of the volume integrals $J_R$ with $A$ is small, whereas
  the preformation factor $P$ increases to lower mass numbers $A$.
  Local maxima for $P$ can be found around 
  $A = 144$ and $A = 208$ corresponding to the magic neutron numbers
  $N =82$ and $N = 126$.
  For $^{174}$Hf = $^{170}$Yb $\otimes$ $\alpha$ the value derived from
  $E = 2.584$\,MeV \protect\cite{Buck93} is shown; the value derived from
  $E = 2.4948$\,MeV \protect\cite{Audi95} exceeds the scale of the diagram
  and is indicated by an arrow (see text and Table \protect\ref{tab:res}).
}
\end{figure}

The results of the calculations are summarized in Table \ref{tab:res}
and shown in Fig.~\ref{fig:res}. One important result is that the 
strength parameters $\lambda$ and the volume 
integrals $J_R$ for all $\alpha$ emitters show only small variations
over the analyzed mass region $A \ge 140$. 
This means that the $\alpha$-nucleus
potential is well defined at very low energies. 
As expected from the
systematic study in \cite{Atz96}, for the light system
$^8$Be = $^4$He $\otimes$ $\alpha$ a much higher
volume integral is required.

The preformation factors $P$ systematically increase to smaller
masses with local maxima around the magic neutron numbers
$N = 82$ and $N = 126$. The very high value of $P = 65\,\%$
for $^8$Be is not surprising because of the well-known
$\alpha$ cluster structure of this nucleus. 

One further exception is
found for $^{174}$Hf = $^{170}$Yb $\otimes$ $\alpha$ with $P = 62.8\,\%$.
However, this surprisingly large value reduces to $P = 3.13\,\%$ if the
energy $E = 2.584$\,MeV from \cite{Buck93} is taken 
which was derived from the
measured $\alpha$ energy and corrected for recoil and atomic effects
instead of $E = 2.4948$\,MeV from \cite{Audi95}.
On the other hand, the uncertainty of the measured half-life 
of $^{174}$Hf is also 20\,\%, and a previous experiment gives
a half-life which is roughly a factor of 2 higher \cite{NNDC}.
From these calculations I have strong evidence that there is an
inconsistency in the system $^{174}$Hf = $^{170}$Yb $\otimes$ $\alpha$
between the measured half-life, the $\alpha$ energy,
and the masses from the mass table \cite{Audi95}.

\section{Conclusions and Outlook}
\label{sec:out}
\begin{figure}
\begin{center}
\epsfig{figure=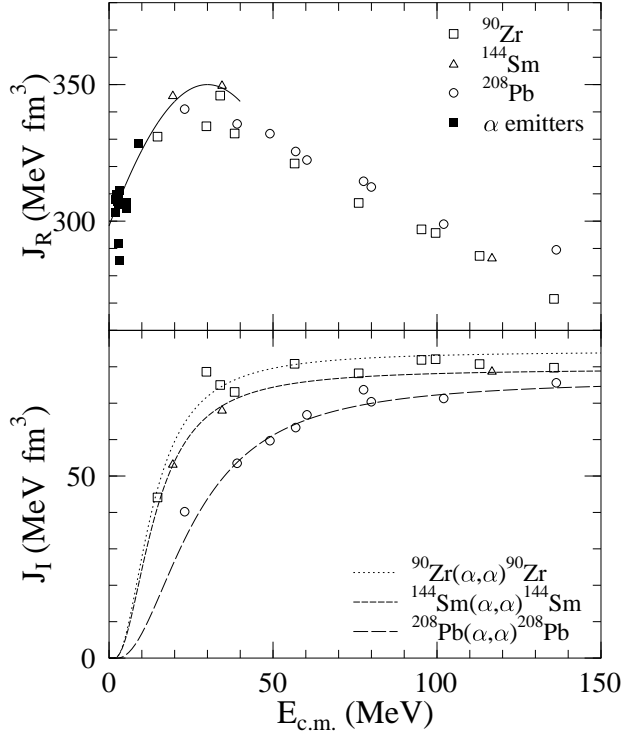,bbllx=60,bblly=60,bburx=580,bbury=580,
width=10.0cm}
\end{center}
\caption{
  \label{fig:vol} 
  Volume integrals $J_R$ (upper diagram) 
  derived from $\alpha$ decay data (this work)
  and from elastic scattering \protect\cite{Atz96,Mohr97}
  for the nuclei $^{90}$Zr, $^{144}$Sm, and $^{208}$Pb.
  The full line shows the recommended interpolation for the
  astrophysically relevant energy region.
  The lower diagram shows the volume integral $J_I$ of the
  imaginary part of the potential for the same nuclei
  together with Brown-Rho parametrizations \protect\cite{Brown81}
  (dashed and dotted lines). 
  This figure is
  combined from data of \protect\cite{Atz96,Mohr97}.
}
\end{figure}

The real part of the optical potential is well defined by the systematic
study of scattering data at higher energies above about 20\,MeV
and by the adjustment of the potential to the bound state properties
at very low energies (see Fig.~\ref{fig:vol}). 
An interpolation between these energy regions
leads to the recommended volume integral which is shown in Fig.~\ref{fig:vol}
as full line. A Gaussian parametrization is applied to $J_R$:
\begin{equation}
J_R(E) = J_{R,0} \times \exp{[-(E-E_0)^2/\Delta^2]}
\label{eq:jr}
\end{equation}
with $J_{R,0} = 350\,{\rm{MeV\,fm^3}}$, $E_0 = 30$\,MeV, and
$\Delta = 75$\,MeV. This interpolation is valid from very low
energies up to about 40\,MeV.

The energy dependence of $J_R$ at low energies is 
$\Delta J_R/\Delta E \approx 1.7$\,MeVfm$^3$/MeV, in agreement with
$\Delta J_R/\Delta E = 1 - 2$\,MeVfm$^3$/MeV \cite{Mohr97} and
somewhat larger than
$\Delta J_R/\Delta E = 0.71$\,MeVfm$^3$/MeV \cite{Gra98}.
The uncertainty of an interpolated value of $J_R$ is significantly
smaller than 10\,MeV\,fm$^3$ corresponding to about 3\,\%
in the interesting energy range around 10\,MeV.

Whereas the real part of the potential is well defined, no experimental
information is available for the imaginary part of the potential
at very low energies. However, 
it has been shown that transmission coefficients and cross sections
in statistical model calculations depend sensitively on the
volume integral $J_I$ and even the shape of the imaginary part of
the potential. Fig.~\ref{fig:vol} shows also the volume integrals $J_I$
of the imaginary part of the potential for the same nuclei
together with a parametrization
from \cite{Brown81}. A different parametrization of the energy dependence
of $J_I$ has been presented in \cite{Gra98}.
Because of the very similar behavior of $^{90}$Zr
and $^{144}$Sm further information could be obtained from $\alpha$ scattering
in the $A \approx 100$ region where the energy region between 10 and 20\,MeV
might be accessible for high-precision scattering experiments. 
Note that in the
$^{144}$Sm case an analysis of scattering data below $E = 20$\,MeV
\cite{Mohr97} is very difficult because of the dominating Coulomb
interaction.

New experiments in the $A \approx 100$ region are planned. If carried out
with sufficient precision, the predictions of this work for the real
part of the potential can be tested and new information on the
imaginary part can be derived.

\acknowledgements 
Discussions with H.\ Oberhummer, E.\ Somorjai, G.\ Staudt, and A.\ Zilges
are gratefully acknowledged.

\begin{table}
  \caption{\label{tab:res} 
    Experimental and theoretical decay widths
    of several neutron-deficient $\alpha$ emitters (parent).
    The light $\alpha$ emitter $^8$Be has been added.
}
\begin{center}
\begin{tabular}{cccccccr@{$\pm$}lc}
parent
& daughter 
& $E\,({\rm{MeV}})$ 
& $Q$
& $\lambda$ 
& $J_R\,({\rm{MeV\,fm}}^3)$
& $T_{1/2}^{\rm{calc}}\,({\rm{y}})$
& \multicolumn{2}{c}{$T_{1/2}^{\rm{exp}}\,({\rm{y}})$}
& $P$ (\%) \\
\hline
$^{212}$Po & $^{208}$Pb & 8.9541 & 22 & 1.241 & 328.4 
& $9.20 \times 10^{-9}\,{\rm{s}}$
& $(2.99$
& $0.02) \times 10^{-7}\,{\rm{s}}$
& 3.08 \tablenotemark[1] \\
$^{210}$Po & $^{206}$Pb & 5.4075 & 20 & 1.148 & 304.5 
& $2.35 \times 10^{-3}$
& $0.37886$
& $0.00001$
& 0.62 \\
$^{208}$Po & $^{204}$Pb & 5215.5 & 20 & 1.157 & 306.8 
& $3.27 \times 10^{-2}$
& $2.898$
& $0.002$
& 1.13 \\
$^{190}$Pt & $^{186}$Os & 3.2495 & 18 & 1.067 & 285.5 
& $3.01 \times 10^{10}$
& $(6.5$
& $0.3) \times 10^{11}$
& 4.63 \\
$^{186}$Os & $^{182}$W  & 2.8220 & 18 & 1.078 & 291.9 
& $9.08 \times 10^{13}$
& $(2.0$
& $1.1) \times 10^{15}$
& 4.54 \\
$^{174}$Hf & $^{170}$Yb & 2.4948 & 18 & 1.116 & 309.3
& $1.26 \times 10^{15}$
& $(2.0$
& $0.4) \times 10^{15}$
& 62.8 \tablenotemark[2] \\
$^{154}$Dy & $^{150}$Gd & 2.9466 & 18 & 1.125 & 309.1 
& $1.16 \times 10^{5}$
& $(3.0$
& $1.5) \times 10^{6}$
& 3.86 \\
$^{152}$Gd & $^{148}$Sm & 2.2046 & 18 & 1.146 & 309.7 
& $1.42 \times 10^{13}$
& $(1.08$
& $0.08) \times 10^{14}$
& 13.1 \\
$^{150}$Gd & $^{146}$Sm & 2.8089 & 18 & 1.141 & 306.1 
& $1.73 \times 10^{5}$
& $(1.79$
& $0.08) \times 10^{6}$
& 9.65 \\
$^{148}$Gd & $^{144}$Sm & 3.2712 & 18 & 1.159 & 311.2 
& $6.20$
& $74.6$
& $3.0$
& 8.31 \\
$^{148}$Sm & $^{144}$Nd & 1.9860 & 18 & 1.123 & 303.2 
& $1.08 \times 10^{15}$
& $(8$
& $2) \times 10^{15}$
& 13.5 \\
$^{146}$Sm & $^{142}$Nd & 2.5289 & 18 & 1.138 & 307.1 
& $9.53 \times 10^{6}$
& $(1.03$
& $0.05) \times 10^{8}$
& 9.25 \\
$^{144}$Nd & $^{140}$Ce & 1.9052 & 18 & 1.147 & 307.9 
& $4.58 \times 10^{14}$
& $(2.29$
& $0.16) \times 10^{15}$
& 20.0 \\
$^{8}$Be & $^{4}$He & 0.0919 & 4 & 1.624 & 444.2
& $4.34 \times 10^{-17}\,{\rm{s}}$
& $(6.71$
& $1.68) \times 10^{-17}\,{\rm{s}}$
& 64.6 \\
\end{tabular}
\tablenotetext[1]{
  Minor differences between the preformation factor $P$
  in Ref.~\protect\cite{Hoy94} and this work are due to the different choice
  of the Coulomb radius: 
  $R_C = 1.2\,{\rm{fm}} \times A_T^{1/3} = 7.11\,{\rm{fm}}$ 
  \protect\cite{Hoy94}
  and $R_C = 6.099\,{\rm{fm}} = r_{rms}$ (this work).
}
\tablenotetext[2]{
  A value of $P = 3.13$\,\% 
  is achieved with the energy 
  $E = 2.584$\,MeV
  from Ref.~\protect\cite{Buck93} (see text).
}
\end{center}
\end{table}


\end{document}